\documentclass[aps,prl,twocolumn,amssymb,amsmath,amssymb,amsfonts,superscriptaddress,showpacs]{revtex4-1}
\usepackage{ulem}
\usepackage{color}
\usepackage{graphicx,graphics,pstricks,epsfig,wrapfig}

 \usepackage{graphicx}

\begin{document}


\title{Total Absorption Spectroscopy Study of $^{92}$Rb Decay: A Major Contributor to Reactor Antineutrino Spectrum Shape}


\author{A.-A. Zakari-Issoufou}
\affiliation{SUBATECH, CNRS/IN2P3, Universit\'e de Nantes, Ecole des Mines de Nantes, F-44307 Nantes, France}
\author {M. Fallot}
\affiliation{SUBATECH, CNRS/IN2P3, Universit\'e de Nantes, Ecole des Mines de Nantes, F-44307 Nantes, France}
\author{ A. Porta}
\affiliation{SUBATECH, CNRS/IN2P3, Universit\'e de Nantes, Ecole des Mines de Nantes, F-44307 Nantes, France}
\author{ A. Algora}
\affiliation{Instituto de Física Corpuscular (CSIC-Universitat de Valencia), Apartado Correos 22085, E-46071 Valencia, Spain}
\affiliation{Institute of Nuclear Research, MTA ATOMKI, Debrecen, 4026 Hungary}
\author{ J.L. Tain}
\affiliation{Instituto de Física Corpuscular (CSIC-Universitat de Valencia), Apartado Correos 22085, E-46071 Valencia, Spain}
\author{ E. Valencia}
\affiliation{Instituto de Física Corpuscular (CSIC-Universitat de Valencia), Apartado Correos 22085, E-46071 Valencia, Spain}
\author{ S. Rice}
\affiliation{Department of Physics, University of Surrey, Guildford GU27XH, United Kingdom}
\author{ V.M Bui}
\affiliation{SUBATECH, CNRS/IN2P3, Universit\'e de Nantes, Ecole des Mines de Nantes, F-44307 Nantes, France}
\author{ S. Cormon}
\affiliation{SUBATECH, CNRS/IN2P3, Universit\'e de Nantes, Ecole des Mines de Nantes, F-44307 Nantes, France}
\author{ M. Estienne}
\affiliation{SUBATECH, CNRS/IN2P3, Universit\'e de Nantes, Ecole des Mines de Nantes, F-44307 Nantes, France}
\author{ J. Agramunt}
\affiliation{Instituto de Física Corpuscular (CSIC-Universitat de Valencia), Apartado Correos 22085, E-46071 Valencia, Spain}
\author{ J. \"Ayst\"o}
\affiliation{Helsinki Institute of Physics, University of Helsinki, FI-00014 Helsinki, Finland}
\author{ M. Bowry}
\affiliation{Department of Physics, University of Surrey, Guildford GU27XH, United Kingdom}
\author{ J.A. Briz}
\affiliation{SUBATECH, CNRS/IN2P3, Universit\'e de Nantes, Ecole des Mines de Nantes, F-44307 Nantes, France}
\author{ R. Caballero-Folch}
\affiliation{Universitat Polit\'ecnica de Catalunya (UPC), 08034 Barcelona, Spain}
\author{ D. Cano-Ott}
\affiliation{Centro de Investigaciones Energéticas Medioambientales Y Tecnológicas, E-28040 Madrid, Spain}
\author{ A. Cucoanes}
\affiliation{SUBATECH, CNRS/IN2P3, Universit\'e de Nantes, Ecole des Mines de Nantes, F-44307 Nantes, France}
\author{ V.-V. Elomaa}
\affiliation{Department of Physics, University of Jyv\"askyl\"a, P.O. Box 35, FI-40014 Jyv\"askyl\"a, Finland}
\author{ T.  Eronen}
\affiliation{Department of Physics, University of Jyv\"askyl\"a, P.O. Box 35, FI-40014 Jyv\"askyl\"a, Finland}
\author{ E. Est\'evez}
\affiliation{Instituto de Física Corpuscular (CSIC-Universitat de Valencia), Apartado Correos 22085, E-46071 Valencia, Spain}
\author{ G.F. Farrelly}
\affiliation{Department of Physics, University of Surrey, Guildford GU27XH, United Kingdom}
\author{ A.R. Garcia}
\affiliation{Centro de Investigaciones Energéticas Medioambientales Y Tecnológicas, E-28040 Madrid, Spain}
\author{ W. Gelletly}
\affiliation{Instituto de Física Corpuscular (CSIC-Universitat de Valencia), Apartado Correos 22085, E-46071 Valencia, Spain}
\affiliation{Department of Physics, University of Surrey, Guildford GU27XH, United Kingdom}
\author{ M.B Gomez-Hornillos}
\affiliation{Universitat Polit\'ecnica de Catalunya (UPC), 08034 Barcelona, Spain}
\author{ V. Gorlychev}
\affiliation{Universitat Polit\'ecnica de Catalunya (UPC), 08034 Barcelona, Spain}
\author{ J. Hakala}
\affiliation{Department of Physics, University of Jyv\"askyl\"a, P.O. Box 35, FI-40014 Jyv\"askyl\"a, Finland}
\author{ A. Jokinen}
\affiliation{Department of Physics, University of Jyv\"askyl\"a, P.O. Box 35, FI-40014 Jyv\"askyl\"a, Finland}
\author{ M.D. Jordan}
\affiliation{Instituto de Física Corpuscular (CSIC-Universitat de Valencia), Apartado Correos 22085, E-46071 Valencia, Spain}
\author{ A. Kankainen}
\affiliation{Department of Physics, University of Jyv\"askyl\"a, P.O. Box 35, FI-40014 Jyv\"askyl\"a, Finland}
\author{ P. Karvonen}
\affiliation{Department of Physics, University of Jyv\"askyl\"a, P.O. Box 35, FI-40014 Jyv\"askyl\"a, Finland}
\author{ V.S. Kolhinen}
\affiliation{Department of Physics, University of Jyv\"askyl\"a, P.O. Box 35, FI-40014 Jyv\"askyl\"a, Finland}
\author{ F.G Kondev}
\affiliation{Argonne National Laboratory, Argonne, IL 60439, USA}
\author{ T. Martinez}
\affiliation{Centro de Investigaciones Energéticas Medioambientales Y Tecnológicas, E-28040 Madrid, Spain}
\author{ E. Mendoza}
\affiliation{Centro de Investigaciones Energéticas Medioambientales Y Tecnológicas, E-28040 Madrid, Spain}
\author{ F. Molina}
\affiliation{Instituto de Física Corpuscular (CSIC-Universitat de Valencia), Apartado Correos 22085, E-46071 Valencia, Spain}
\affiliation{Present address: Comisi\'{o}n Chilena de Energ\'{i}a Nuclear, Post Office Box 188-D, Santiago, Chile}
\author{ I. Moore}
\affiliation{Department of Physics, University of Jyv\"askyl\"a, P.O. Box 35, FI-40014 Jyv\"askyl\"a, Finland}
\author{ A. B. Perez-Cerd\'an}
\affiliation{Instituto de Física Corpuscular (CSIC-Universitat de Valencia), Apartado Correos 22085, E-46071 Valencia, Spain}
\author{ Zs. Podoly\'ak}
\affiliation{Department of Physics, University of Surrey, Guildford GU27XH, United Kingdom}
\author{ H. Penttil\"a}
\affiliation{Department of Physics, University of Jyv\"askyl\"a, P.O. Box 35, FI-40014 Jyv\"askyl\"a, Finland}
\author{ P.H. Regan}
\affiliation{Department of Physics, University of Surrey, Guildford GU27XH, United Kingdom}
\affiliation{National Physical Laboratory, Teddington, Middlesex, TW11 0LW, United Kingdom}
\author{ M. Reponen}
\affiliation{Department of Physics, University of Jyv\"askyl\"a, P.O. Box 35, FI-40014 Jyv\"askyl\"a, Finland}
\affiliation{Present address: RIKEN, 2-1 Hirosawa, Wako, Saitama 351-0198, Japan}
\author{ J. Rissanen}
\affiliation{Department of Physics, University of Jyv\"askyl\"a, P.O. Box 35, FI-40014 Jyv\"askyl\"a, Finland}
\author{ B. Rubio}
\affiliation{Instituto de Física Corpuscular (CSIC-Universitat de Valencia), Apartado Correos 22085, E-46071 Valencia, Spain}
\author{ T. Shiba}
\affiliation{SUBATECH, CNRS/IN2P3, Universit\'e de Nantes, Ecole des Mines de Nantes, F-44307 Nantes, France}
\author{ A.A. Sonzogni}
\affiliation{National Nuclear Data Center, Brookhaven National Laboratory, Upton NY 11973-5000, USA}
\author{ C. Weber}
\affiliation{Department of Physics, University of Jyv\"askyl\"a, P.O. Box 35, FI-40014 Jyv\"askyl\"a, Finland}
\affiliation{Present address: Faculty of Physics, Ludwig-Maximilians University Munich, Am Coulombwall 1, D-85748 Garching, Germany}
\author{ IGISOL Collaboration}
\affiliation{Department of Physics, University of Jyv\"askyl\"a, P.O. Box 35, FI-40014 Jyv\"askyl\"a, Finland}


\date{\today}

\begin{abstract}
The antineutrino spectra measured in recent experiments at reactors are inconsistent with calculations based on the conversion of integral beta spectra recorded at the ILL reactor. $^{92}$Rb makes the dominant contribution to the reactor antineutrino spectrum in the 5-8 MeV range but its decay properties are in question. We have studied $^{92}$Rb decay with total absorption spectroscopy. Previously unobserved beta feeding was seen in the 4.5-5.5 region and the GS to GS feeding was found to be 87.5(25)$\%$. The impact on the reactor antineutrino spectra calculated with the summation method is shown and discussed.
\end{abstract}


\maketitle



Beta decay properties of fission products are at the origin of the antineutrino flux emitted by reactor cores. This flux has been used for decades as a source for reactor neutrino experiments, such as Daya Bay, Double Chooz and Reno which have recently published their new results for the mixing angle $\theta_{13}$\,\cite{DC,daya,reno}. These results will allow future searches for the CP violation phase $\delta$ or the neutrino mass hierarchy with complementary experiments at reactors\,\cite{juno}. 
The accurate determination and understanding of the emitted reactor antineutrino flux is thus still required for present and future experiments. The recent re-estimate of reactor antineutrino energy spectra\,\cite{Mueller, Huber} has led to the so-called ``reactor anomaly"\,\cite{anomaly}, at the origin of new experimental projects chasing short distance oscillations at research reactors~\cite{Sterile}. These calculations are based on the conversion into antineutrinos of the only available measurement of the beta energy spectra performed using the high flux reactor at the Institut Laue-Langevin (ILL) in Grenoble, France ~\cite{Schreck}. The conversion method has until now been considered as the most precise one by experimenters studying neutrino oscillations. But recently, Hayes et al.\,\cite{Hayes} have shown that it is dependent on the underlying nuclear physics and that the associated errors should be revised.
In addition, very recent experimental results from\,\cite{DC,daya,reno} have shown an unexplained distortion between 4 and 8\,MeV in their measured positron energy spectra from Pressurized Water Reactors (PWR)\,\cite{bump} with respect to the converted spectra\,\cite{Mueller, Huber} (an excess of ca. 10\% over 2\,MeV followed by a dip). The positron energy used in reactor antineutrino experiments corresponds to the antineutrino energy minus the mass difference between the neutron and proton. In this context, new evaluations of PWR antineutrino energy spectrum are essential. 

An alternative method, independent of the ILL measurements, relies on the summation of the contributions of the fission product beta decay branches to obtain the antineutrino energy spectra. The need to measure new nuclear physics properties of some major contributors to the antineutrino spectrum was underlined in\,\cite{Fallot}, where it was shown that we should use the Total Absorption Spectroscopy (TAS) technique to avoid the pandemonium effect\,\cite{pandemonium} and improve the predictions of the summation method. The summation method is indeed the only one which allows the prediction of antineutrino spectra for which no integral beta measurement exists. This is required, for instance, in the context of the R\&D of antineutrino detection as a tool for reactor monitoring\,\cite{ESARDA}.

In this Letter, we present the first results of an experimental campaign performed with TAS technique \,\cite{ExpProp} aimed at the measurement of beta decay properties of important contributors to the reactor antineutrino energy spectrum emitted by PWRs. In particular, we show new results for $^{92}$Rb, the largest contributor to the reactor antineutrino flux in the energy range above 5\,MeV. In the following, we present a short list of nuclei making the main contributions to the antineutrino energy spectrum above 4\,MeV, obtained using the summation method presented in\,\cite{Fallot}. Then, previous experimental knowledge of the beta decay properties of $^{92}$Rb is summarized, and the TAS method, the experimental setup used and the data analysis performed are presented. Finally, we show the beta feeding obtained and present the impact of the new results on reactor antineutrino energy spectra.


The main contributors to the antineutrino energy spectra from 4 to 8\,MeV are listed in Table\,\ref{tableNucl}. 
In our calculation, we have chosen to minimize the impact of the pandemonium effect on the antineutrino spectra and on the computed proportions of the nuclei per energy bin. For this purpose, we have used data from\,\cite{Rudstam} for $^{92, 93, 94}$Rb, $^{96}$Y, $^{142}$Cs, $^{135}$Te and from\,\cite{Greenwood} for $^{95}$Sr and $^{90}$Rb, because these two sets of data are likely to be pandemonium free (though they may suffer from other systematic errors). As they were not measured by\,\cite{Greenwood} or\,\cite{Rudstam}, data for $^{98m}$Y and $^{100}$Nb were taken from\,\cite{JEFF3.1} and $^{104m}$Nb from\,\cite{ensdfnew}. 
Indeed, a careful choice of data sets is needed, especially to select nuclei which would deserve new measurements, as is illustrated below with the case of $^{92}$Rb.
Note that in the 4 to 6\,MeV range, unknown nuclei requiring the use of models represent less than 1\% of the spectrum, while they represent about 4\% of the 7 to 8\,MeV bin.

\begin{table}[b]
\caption{\label{tableNucl}
Main contributors to a standard PWR antineutrino energy spectrum computed with the MURE code coupled with the list of nuclear data given in\,\cite{Fallot}, assuming that they have been emitted by $^{235}$U (52\%), $^{239}$Pu (33\%), $^{241}$Pu (6\%)and $^{238}$U (8.7\%) for a 450~day irradiation time and using the summation method described in\,\cite{Fallot}.}
\begin{ruledtabular}
\begin{tabular}{ccccc}
\textrm{ }&
\textrm{4 - 5\,MeV}&
\textrm{5 - 6\,MeV}&
\textrm{6 - 7\,MeV}&
\textrm{7 - 8\,MeV} \\
$^{92}$Rb & 4.74\% & 11.49\% & 24.27\% &   37.98\%\\
$^{96}$Y & 5.56\% & 10.75\% & 14.10\% &  - \\
$^{142}$Cs & 3.35\% & 6.02\% & 7.93\% & 3.52\%   \\
$^{100}$Nb & 5.52\% & 6.03\%  & -  & -  \\
$^{93}$Rb & 2.34\% & 4.17\% & 6.78\% &  4.21\% \\
$^{98m}$Y  & 2.43\% & 3.16\% & 4.57\% & 4.95\% \\
$^{135}$Te  & 4.01\% & 3.58\% & - & -  \\ 
$^{104m}$Nb  & 0.72\% & 1.82\% & 4.15\% &  7.76\%  \\
$^{90}$Rb  & 1.90\% & 2.59\% & 1.40\% &  -  \\
$^{95}$Sr  & 2.65\% & 2.96\% & - & -  \\
$^{94}$Rb  & 1.32\% & 2.06\% & 2.84\% & 3.96\%   \\
\end{tabular}
\end{ruledtabular}
\end{table}

$^{92}$Rb makes the main contribution between 4 to 8\,MeV, representing alone up to about 38\,\% of the 7 to 8\,MeV bin and 16\,\% of the 5 to 8\,MeV range. 
$^{92}$Rb is quite controversial: the beta feeding to the ground state of its daughter nucleus, $^{92}$Sr, was fixed at 51$\% \pm 18 \%$ in the ENSDF data base\,\cite{ensdfold} until 2012, before the inclusion in the references of the article from Lhersonneau et al.\,\cite{lhersonneau}, 
which concluded that close to half of the decay intensity, mostly high energy ground state transitions, is missing in the decay scheme. Following this reference 
the beta feeding to the ground state of $^{92}$Sr was recently changed  to 95.2 $\% \pm 0.7 \%$ in the ENSDF database \,\cite{ensdfnew}.

$^{92}$Rb has a large $Q_{\beta}$ value which makes it a good candidate to be a pandemonium nucleus. The pandemonium effect\,\cite{pandemonium} arises from the difficulty encountered in building level schemes for complex beta decays using Germanium detectors, especially when beta transitions occur to high-energy levels or regions of high level density. This leads to an underestimate of the corresponding beta branches to states at high excitation energy and thus to a distortion in the beta decay feeding.
In addition, $^{92}$Rb has also been used as a critical example\,\cite{moeller} to show how beta-decay strength calculations impact on the predictive power of models in reconstructing half-lives and beta-delayed neutron emission probabilities of nuclei, whose properties are important in the simulation of the astrophysical r-process. It is also on NEA/IAEA lists of important contributors to reactor decay heat~\cite{NEA-IAEA}.

A total absorption spectrometer is a calorimeter measuring the gamma cascades emitted by the deexcitation of the daughter nucleus after beta decay of the parent. 
The detection of the total energy allows the deduction of the feeding probability of excited levels populated in the beta decay. This quantity is calculated by solving the ``inverse problem'' as presented below. The beta feeding $f$ gives direct access to the beta intensity $I_i = f_i/ \Sigma_k f_k $ and then to the beta strength, a microscopic quantity that can be directly compared with models \,\cite{PRCAlgora}. 
The detector used in the measurement of the $^{92}$Rb decay is composed of 12 crystals of BaF$_{2}$ arranged in a compact geometry described in 
\,\cite{ROCINANTES}. Each crystal is coupled to a photomultiplier tube converting the scintillation light into an electrical signal directly proportional to the detected energy. The gamma detection efficiency was $\sim$ 80\,$\%$ at 5\,MeV.
This spectrometer was coupled to a silicon detector placed in the center, behind the source implantation zone, to tag the beta emission. This reduces the background by demanding coincidences between beta events and the following gamma emission from deexcitation of levels in the daughter.

$^{92}$Rb ions were produced via proton-induced fission on an uranium target at the IGISOL facility \,\cite{jyv} in the accelerator Laboratory of the University of Jyv\"askyl\"a (Finland). JYFLTRAP double Penning trap \,\cite{trap} was used for selecting with high precision only $^{92}$Rb ions using the mass-selective buffer gas cooling technique \,\cite{gascooling}. This high level of purification of the beam is necessary in TAS experiments in order to reduce systematic uncertainties related to the purity of the beam.

As stated above the main observable in a TAS measurement is the beta feeding to the energy levels of the daughter nucleus which is contained in the measured gamma spectrum convoluted with the detector response. To extract this information we have to solve the so-called ``inverse problem''. It consists of solving the equation $d_i=R_{ij}^{(B)} \times f_j$, where $R_{ij}^{(B)}$ is the response matrix of the detector to an assumed decay level scheme $(B)$. $R_{ij}$ connects feeding to level j ($f_j$) to counts in the bin $i$ of the ``measured" TAS spectrum ($d_i$). The analysis procedure has been described in previous publications~\cite{PRLAlgora, PRCAlgora} and is well understood.

To perform the analysis, very clean decay data $d$ are required.
Possible contamination from the daughter nuclei decay and pileup signals are subtracted from the raw data. The shape of the pile-up spectrum has been computed by summing in the ADC time window two events randomly extracted from the raw data. The absolute normalization of the pileup was performed using the data counting rate and the ADC time window in which random coincidences can occur \cite{thesezak, pileup}. The shape of the spectrum from the decay of the daughter nucleus $^{92}$Sr  has been simulated using its known level scheme from ENSDF \cite {ensdfnew} and the detector response. The normalization factor has been obtained solving the Bateman equations for the decay of $^{92}$Rb in realistic experimental conditions, i.e. considering the experimental time for implantation and measuring cycles. The contamination from $^{92}$Sr decay represents 0.08~$\%$ of the total $^{92}$Rb data acquired. 
The response matrix $R$ is calculated by simulating the detector response to beta and gamma cascades emitted during the decay with a dedicated GEANT4 \cite{geant} Monte Carlo simulation. The latter has been validated using measurements performed with known sources in order to reproduce the detector response in great detail \cite{thesezak, thesesimon,simu}. 
The ``inverse problem" is solved by using a maximization expectation algorithm based on the Bayes theorem and combined with a $\chi^{2}$ minimization \cite{bayes}. It makes use of an iterative method to find the final feeding distribution by minimizing the difference between the experimental data and the spectrum recreated by the result of the algorithm at each iteration. The analysis starts with a first guess at feeding values extracted from the literature, or an equally probable feeding distribution if the nucleus is poorly known, and stops when the $\chi^{2}$ value deduced from the two spectra is at a minimum.

The starting point for solving the inverse problem is the construction of the branching ratio matrix $(B)$  for the states populated in the decay. For this purpose we begin by using the known information, derived from high resolution studies, about levels up to an excitation energy  of  1778~keV in $^{92}$Sr ~\cite{ensdfnew}. Above this energy little is known and the data are divided into 40 keV bins up to the $Q_{\beta}$ value. In this range we must have recourse to semiempirical statistical models and we must supply as input both the level densities and gamma strength functions. Three level-density models were tested: Back-Shifted-Fermi-Gas (BSFG) \,\cite{Dilg, Egidy}, Constant-Temperature \,\cite{Egidy}, and Gilbert-Cameron models \,\cite{Gilbert}. The last of these is a combination of the other two. The Gilbert-Cameron formulation was chosen because it best reproduces the experimental data at low energies.
 The gamma strengths were modeled with a Lorentz function using the parameters given in \,\cite{RIPL2}. 
 In determining the $\beta$-feeding distribution, it is possible to fix or vary the feeding to each individual level or energy bin. The feeding to the 1673.3 keV level was set to zero, since the probable spin parity is 4$^{+}$ and any feeding from the $^{92}$Rb ground state must be negligible.

\begin{figure}  
\includegraphics[width=8cm]{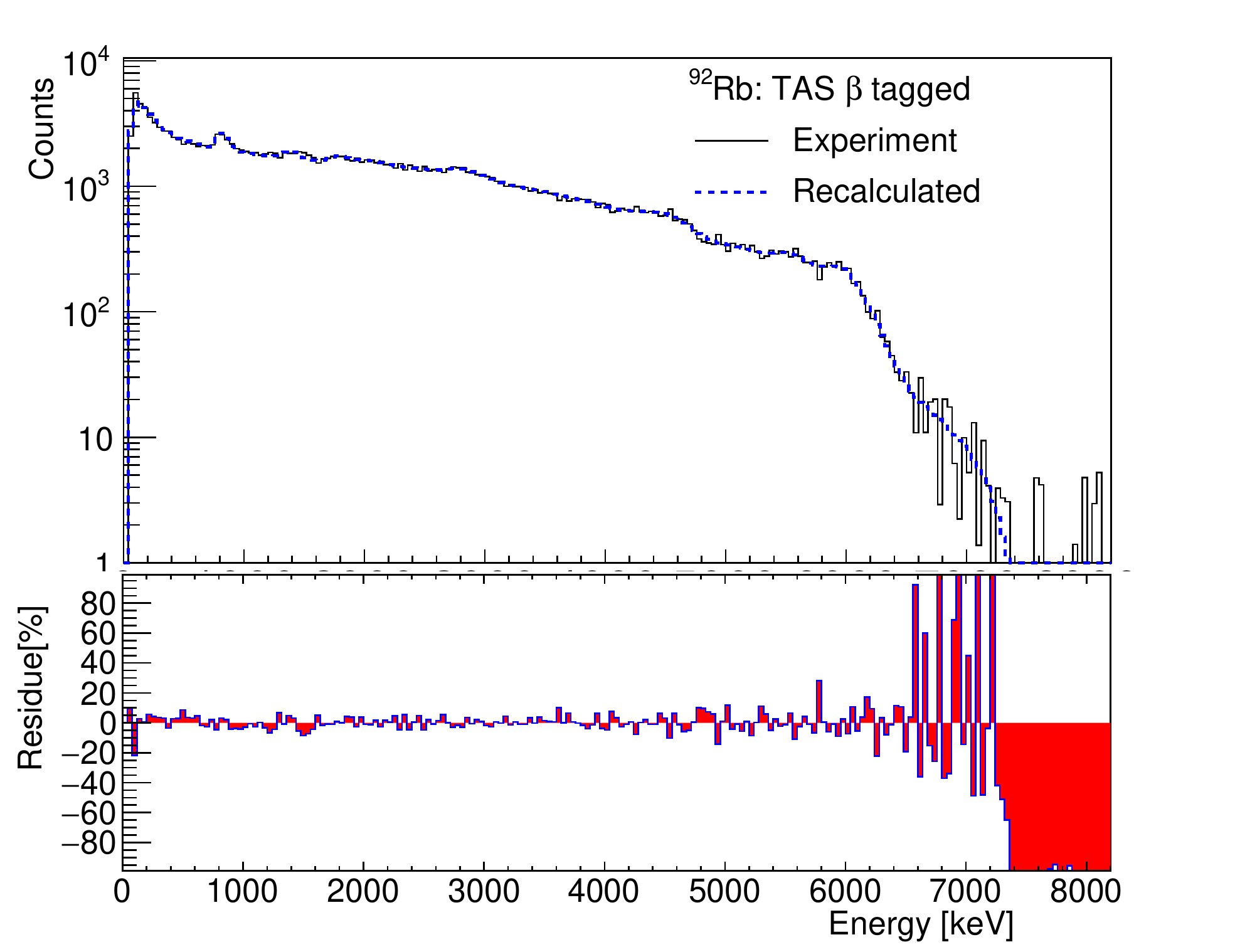}
\caption{Upper panel: Comparison between measured spectrum (black continuous line) and reconstructed one (blue dashed line) with the feeding obtained from the TAS data analysis. Lower panel: Residues between the two curves reported in the upper panel.}
 \label{comp}
 \end{figure}

\begin{figure}
\includegraphics[width=8cm]{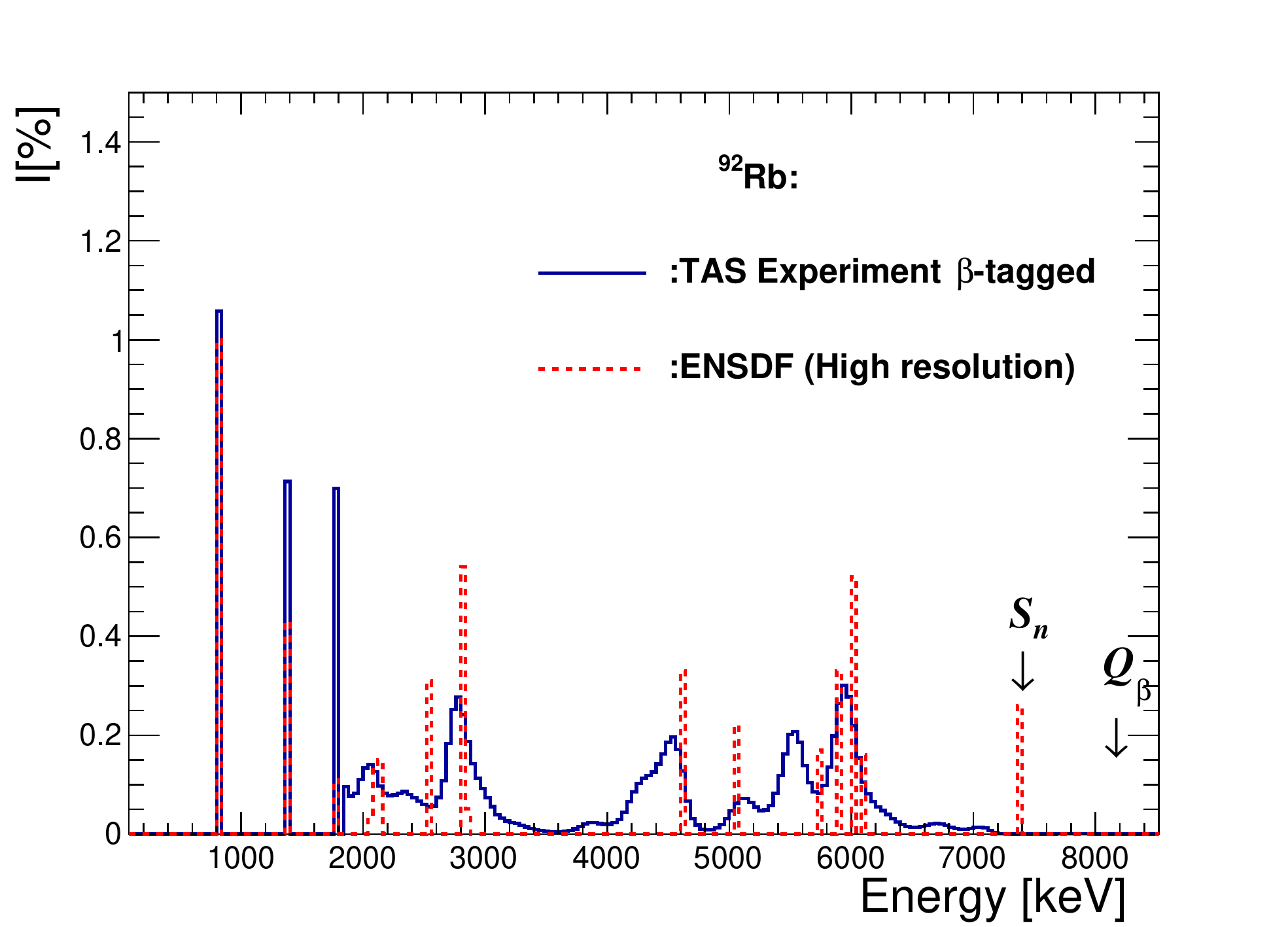}
\caption{Beta intensity for the decay of $^{92}$Rb obtained with TAS measurements in the blue continuous line. Red dashed lines are data from ENSDF ~\cite{ensdfnew}. Ground state feeding from ENSDF is $95.2 \% \pm 0.7 \%$ , while from this work it is $ 87.5 \% \pm 2.5 \%$}
 \label{feed}
 \end{figure}

The reconstructed spectrum (blue dashed line) calculated using the feeding distribution obtained from this analysis is compared with the clean decay data (black continuous line) of $^{92}$Rb in the upper panel of figure\,\ref{comp} \,\cite{thesezak}. The lower panel shows the residues between these two curves.
The beta intensity obtained from the solution of the inverse problem for $^{92}$Rb is shown in figure\,\ref{feed} in the blue continuous line, while the red dashed lines are the intensities from ENSDF \cite{ensdfnew}. As the ground state feeding is very important in the case of the decay of $^{92}$Rb, we have estimated the main errors involved in this reconstruction \cite{thesezak}. They are listed as follows: the threshold of the beta spectrum, statistical uncertainty, error induced by pile-up subtraction, errors in the detector energy calibration and resolution used in the calculation of the response matrix $R$ and errors obtained by testing different input parameters for the  calculation of $R$ and inverse problem resolution. 
A sum in quadrature of all the systematic and statistical errors quoted above gives a 2.5~\% error on the ground state feeding. This result is conservative, as we have voluntarily adopted large values of the main errors which are associated with the threshold of the beta spectrum and with the choice of model for the level density.
The TAS results show some beta intensity around 4.5 and 5.5 MeV which was not detected before. The intensity to the ground state obtained from our analysis is 87.5 (25)~$\%$. This value can be obtained from the data analysis because the TAS detector also measures the bremsstrahlung radiation from the beta particles. These events are in the low energy part of the measured spectrum and, since they are considered in the response matrix $R$, they contribute to the reconstruction of the spectrum and, then, in the calculation of beta feeding. The selected ground state feeding is the one which minimizes the $\chi^2$ value determined from the experimental data and the reconstructed spectrum after the analysis.
If we fix the ground state feeding to be 95.2 $\%$ as reported in the ENSDF data base ~\cite{ensdfnew} our analysis converges with a $\chi^2$ value of 2048 which is much larger than the minimum of 630, completely excluding this hypothesis.
 
The antineutrino energy spectrum emitted in $^{92}$Rb beta decay has been computed using the beta feeding presented above. The GS to GS transition is first-forbidden nonunique (spin parity of $^{92}$Rb 0$^-$). Different spectral shapes were assumed for this transition, considering the different possibilities listed in\,\cite{Hayes}; an allowed shape, a first forbidden nonunique shape due to the GT operator and a first forbidden nonunique shape due to the $\rho_A$ operator. It was also assumed that the remaining transitions were of allowed or first forbidden unique type. The various combinations of these options were computed and the shapes obtained were very similar. No significant impact is expected from the uncertainty of the shape of the first forbidden nonunique GS to GS transition. We chose to adopt an allowed shape for the GS to GS transition and first forbidden unique shapes for the remaining branches due to the spins and parities of the known transitions in this nucleus.

\begin{figure}  
 \includegraphics[width=8cm]{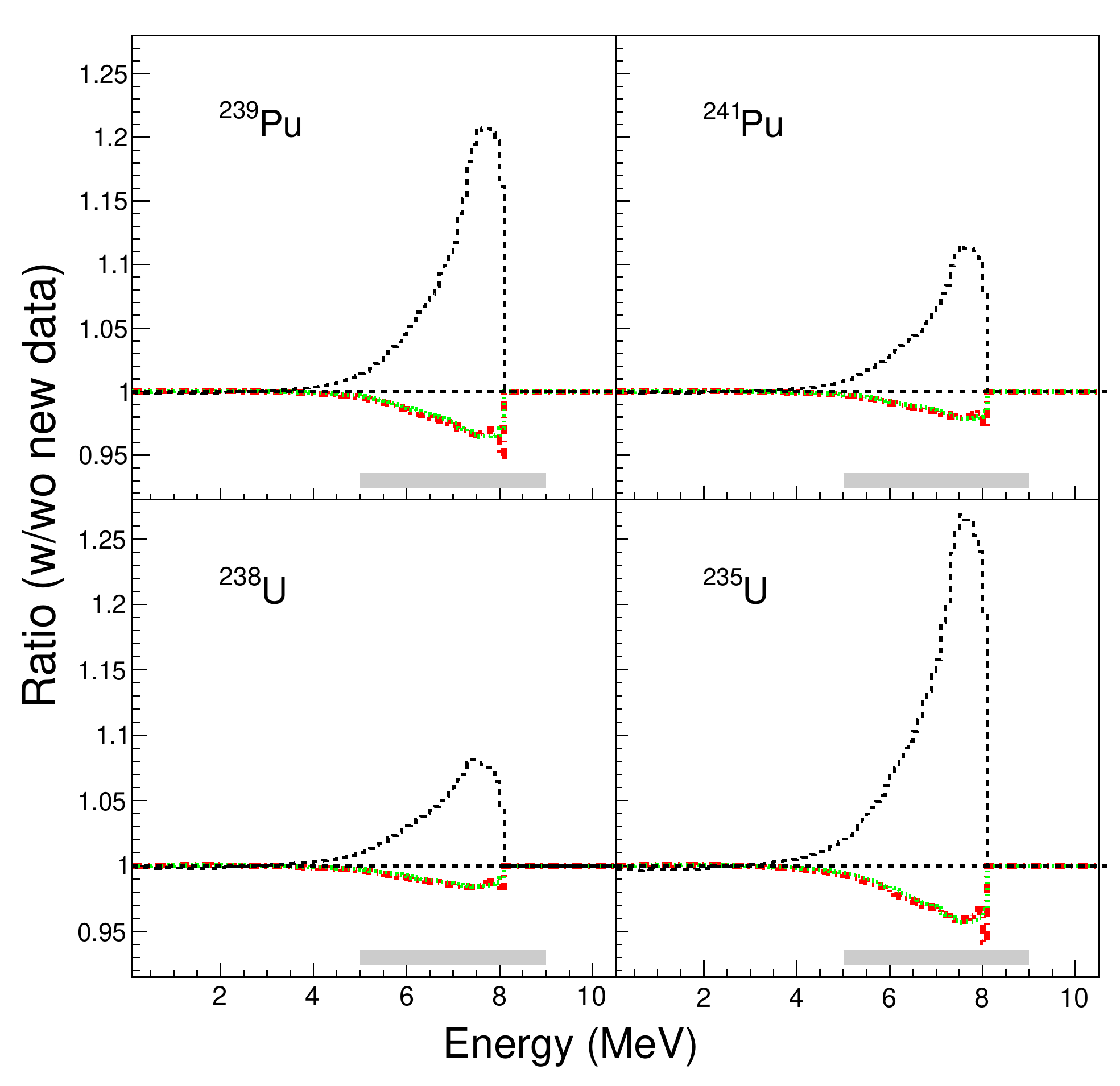}
\caption{Ratio between the antineutrino spectra calculated using the results presented in this paper with respect to the data on $^{92}$Rb decay used in\,\cite{Fallot} (thick red dashed-dotted line), in\,\cite{Sonzogni} (green dotted line) and in\,\cite{Dwyer} (black dashed line). The sharp drop in  the ratio, in one single bin located at the Q value of the $^{92}$Rb, is due to the different values in Q given in\,\cite{Rudstam} and\,\cite{ensdfnew}, that were used to reconstruct the antineutrino spectrum. A gray horizontal bar is placed above the antineutrino energy scale to indicate the region of the distortion observed by the reactor antineutrino experiments whit respect to converted spectra.}
 \label{spectra}
 \end{figure}

The  antineutrino energy spectra were calculated with the summation method described in\,\cite{Fallot}. 
In\,\cite{Fallot} the data adopted for $^{92}$Rb were extracted from\,\cite{Rudstam}. 
In principle, these measurements should not suffer from the pandemonium effect, nor from a lack of knowledge of the types of the beta transitions. Unfortunately, however, the error bars are quite large.
In figure\,\ref{spectra}, the ratio between the antineutrino spectra of $^{239,241}$Pu and $^{235,238}$U from\,\cite{Fallot} and those obtained using our new results for $^{92}$Rb is displayed with the red dashed-dotted line. As expected, the main effect is in the 4 to 8\,MeV antineutrino energy range, with a maximum between 7 and 8\,MeV, and amounts to 4.5$\%$ for $^{235}$U, 3.5$\%$ for $^{239}$Pu, 2$\%$ for $^{241}$Pu and 1.5$\%$ for $^{238}$U. 
These discrepancies are due to the difference in the shapes of the antineutrino spectra built with the newly measured beta feedings with respect to the antineutrino spectra converted from Rudstam's measurements.
The comparison would be very similar if we had used the latest ENSDF\,\cite{ensdfnew} data for $^{92}$Rb in our summation calculations, as was done in\,\cite{Sonzogni}. The ratio is displayed as well in Fig.\,\ref{spectra} with green dotted lines, and is nearly superposed on the ratio built when using Rudstam data in the first place. The change becomes even more dramatic if one compares with summation method spectra in which an older version of the ENSDF data was used, as in\,\cite{Dwyer}. The latter ratio is plotted with black dashed lines in figure\,\ref{spectra}.  
This shows the relevance of the present $^{92}$Rb decay data in the calculations. 

In summary,  the results of new measurements of the beta decay properties of $^{92}$Rb have been presented. This nucleus makes one of the largest contributions to the emitted antineutrino flux by standard thermal reactors in the energy region above 5\,MeV. The measurements have been performed using pure isotopic beams and the TAS technique to provide data free from the pandemonium effect. The measured feeding distribution, which extends to states previously not seen in high-resolution measurements and also determines the GS to GS feeeding as 87.5(25)\%, confirms the relevance of this decay to antineutrino summation calculations. The impact of the measurements has been evaluated by comparing the ratio of summation calculations using the new feeding distribution with the results using the feeding distributions employed in\,\cite{Fallot, Sonzogni, Dwyer}. The effect of introducing the new results is particularly marked in the case of\,\cite{Dwyer} and calls for a revision of the conclusions drawn in that paper. It is clear that this is because the GS to GS feeding used in\,\cite{Dwyer} was incorrect. The overall agreement of the new summation calculations with the converted spectra\,\cite{Huber} is improved in the 4 to 8~MeV range except in the case of $^{235}$U for which the summation method spectrum is always below the converted spectrum. The change is especially striking in the case of \cite{Dwyer} in the 5 to 8~MeV antineutrino energy range, which overlaps the energy region in which reactor neutrino experiments have shown a spectral distortion\,\cite{bump}. 
It also shows that the inclusion of all existing TAS nuclear data (ca.\,37 nuclei) in \cite{Dwyer}'s calculation may change dramatically the spectral shape they compute. Overall, this emphasizes why new measurements are needed for the radioactive decays of importance in the reactor antineutrino spectrum and, in particular, why measurements should be performed with the total absorption method. The present measurement, which reduces significantly the uncertainties associated with the antineutrino summation calculations in the 4 to 8~MeV range, is an important step towards better predictions with the summation method. Provided that in the 4 to 6\,MeV range, unknown nuclei requiring the use of models represent less than 1\% of the spectrum in the summation calculation from\,\cite{Fallot}, one can thus expect a dramatic reduction of the final uncertainty in this range as a long term result of the TAS campaign. In parallel, the impact of the uncertainties of the fission yields on the antineutrino spectrum needs to be evaluated more accurately.\\ 
\,
\\
This work was supported by the CHANDA European Project, the PICS 05761 between CNRS/in2p3 and CSIC, the GEDEPEON research groupment, the NEEDS challenge and STFC (UK).\\
This work was supported by Spanish Ministerio de Econom\'{\i}a y Competitividad under Grants No. FPA2008-06419, FPA2010-17142 and FPA2011-24553, and CPAN CSD-2007-00042 (Ingenio2010).\\
Work at ANL was supported by the U.S. DOE, Office of Nuclear Physics under Contract No. DE-AC02-06CH11357.

\bibliography{papierRb92_vf_raf_final}

\end{document}